# Anomalous Hall-like Transverse Magnetoresistance in Au thin films on $Y_3Fe_5O_{12}$

Tobias Kosub[1], Saül Vélez[2,†], Juan M. Gomez-Perez[2], Luis E. Hueso[2,3], Jürgen Faßbender[1], Fèlix Casanova[2,3], Denys Makarov[1]

[1] Helmholtz-Zentrum Dresden-Rossendorf e.V., Institute of Ion Beam Physics and Materials Research, 01328 Dresden, Germany
[2] CIC nanoGUNE, 20018 Donostia-San Sebastian, Basque Country, Spain
†Present address: Department of Materials, ETH Zürich, 8093 Zürich, Switzerland
[3] IKERBASQUE, Basque Foundation for Science, 48013 Bilbao, Basque Country, Spain

**Anomalous Hall-like signals in platinum in contact with magnetic insulators are common observations that could be explained by either proximity magnetization or spin Hall magnetoresistance. In this work, longitudinal and transverse magnetoresistances are measured in a pure gold thin film on the ferrimagnetic insulator $Y_3Fe_5O_{12}$ (Yttrium Iron Garnet, YIG). We show that both the longitudinal and transverse magnetoresistances have quantitatively consistent scaling in YIG/Au and in a YIG/Pt reference system when applying the Spin Hall magnetoresistance framework. No contribution of an anomalous Hall effect due to the magnetic proximity effect is evident.**

Throughout the last few years, systems of magnetic insulators and nonmagnetic metals (MI/NM) have seen significant attention from the spintronics community [1–10]. Such systems generally display Spin Hall magnetoresistance (SMR) [9–19] enabled by the eponymous Spin Hall effect [20,21] of the metal which also offers elegant charge-spin-interconversion. One of the most popular metals in such systems is platinum (Pt) due to its strong spin orbit coupling, large spin Hall angle and its benign chemical properties. Still, a larger variety of well-studied metals is urgently needed in the field of insulator spintronics for two reasons: Firstly, some metals – Pt being a prototypical example – are close to the Stoner criterion for ferromagnetism and can thus show a strong magnetic proximity effect [22]. When Pt becomes a magnetic conductor in this way, its strong SHE becomes an anomalous Hall effect (AHE) [20] that would mirror the magnetization of a nearby MI [23–26]. In contrast, the same phenomenon could be attributed solely due to the transverse SMR [27–29] or the nonlocal AHE [30] creating an ambiguity about the physics of the anomalous Hall-like signal in MI/NM systems. Secondly, studying more diverse systems can probe the applicability limits of the SMR theory [31]. The longitudinal and transverse SMR magnitudes are typically explained in terms of the real and imaginary components of the interfacial spin-mixing conductivity $G_{\uparrow\downarrow} = G_r + iG_i$ [32,33], but higher order effects are already known [34].

Here, we report longitudinal as well as transverse signatures of the spin Hall magnetoresistance for a gold (Au) layer on YIG and compare this to a YIG/Pt reference system. Au shares or exceeds the excellent chemical and electrical properties of Pt. At the same time, static proximity magnetization is not expected for Au [35] and we take care to avoid dynamic proximity magnetization due to thermal spin pumping [Suppl. Information, 1]. As a result, we exclude possible influences from proximity magnetization and its associated anomalous Hall effect. Using empirical data, we *quantitatively* predict the longitudinal and transverse spin Hall magnetoresistance in YIG/Au and confirm their respective magnitudes by measurements. Hence, the key finding of this work is the experimental observation of the transverse SMR in Au.

We prepared a YIG/Au(10nm) and a YIG/Pt(2nm) system using DC sputtering under identical conditions on top of 3.5-μm-thick liquid phase epitaxy YIG/GGG (gallium gadolinium garnet) substrates from Innovent e.V., Jena, Germany. The larger thickness for the Au metal film was chosen to guarantee the continuity of the film [Suppl. Information, 2]. Hall bars 100 μm wide and 800 μm long were patterned by e-beam lithography and Ar ion milling for both systems.

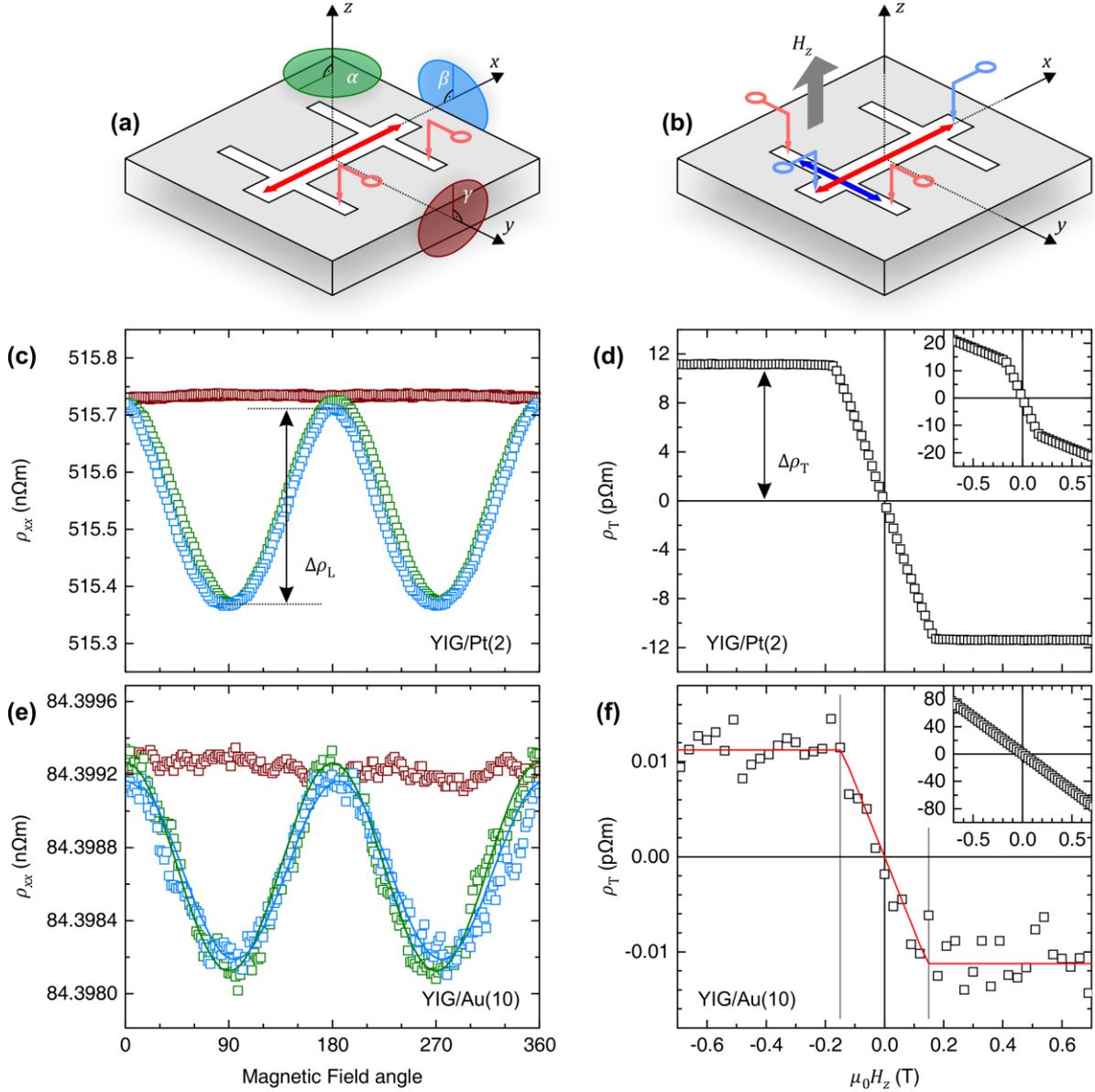

Figure 1: Magnetoresistance measurement geometries: (a) Longitudinal resistivity measurement ($\rho_{xx}$) while rotating the magnetic field along the angles $\alpha$ (green), $\beta$ (blue) and $\gamma$ (brown) and while current is flowing in $x$-direction. (b) Transverse resistivity measurement ($\rho_T$) while sweeping the magnetic field along the $z$-direction and current alternatingly in the $x$- and $y$-directions (Zero-Offset Hall measurement). (c,e) Longitudinal ADMR for YIG/Pt and YIG/Au, respectively, at a magnetic field strength of 1 T. (d,f) Zero-Offset Hall measurement after subtracting the normal Hall effect for YIG/Pt and YIG/Au, respectively. Grey lines indicate the saturation of the YIG and the red line is a data fit. The insets show the data before the subtraction of the normal Hall signal.

The longitudinal resistance in $x$-direction $\rho_{xx}$ and its magnetic field angle dependent magnetoresistance (ADMR) were measured using a conventional Kelvin contact layout [Figure 1(a)] in a $l$-He cryostat with 9 T field and 360° sample rotation capabilities. The transverse resistance and its magnetoresistance were obtained by out-of-plane field Zero-Offset Hall measurements [Figure 1(b)] using an integrated measurement device from HZDR innovations GmbH [more details in Suppl. Information, 2].

In the context of SMR, the resistivity tensor of the NM layer in lab coordinates is given by [12]:

$$\begin{aligned}\rho_{xx} &= \rho_0 - \Delta\rho_1\, m_y^2 \\ \rho_{yy} &= \rho_0 - \Delta\rho_1\, m_x^2 \\ \rho_{xy} &= \Delta\rho_1 m_x\, m_y - \Delta\rho_2\, m_z\end{aligned} \quad (1)$$

where $\mathbf{m}(m_x, m_y, m_z) = \mathbf{M}/M_s$ are the normalized projections of the magnetization of the YIG film the three main axes and $M_s$ is the saturation magnetization of the MI layer. $\rho_0$ is the Drude resistivity and $\Delta\rho_1$ and $\Delta\rho_2$ are the characteristic amplitudes of the SMR. The first term in the expression for $\rho_{xy}$ is a planar Hall effect resulting from the anisotropy of the longitudinal resistivity ($\rho_{\text{PHE}} = \rho_{xx} - \rho_{yy}$). In the following, we do neither discuss nor measure planar Hall effects, which are universally rejected by the Zero-Offset Hall measurement technique [36] due to their origin in the longitudinal tensor components [Suppl. Information, 3]. In our measurement, the transverse signature of the SMR thus simplifies to

$$\rho_{\text{T}} = -\Delta\rho_2\, m_z \quad (2)$$

The longitudinal and transverse amplitudes of the SMR, $\Delta\rho_{\text{L}} \equiv \Delta\rho_1$ and $\Delta\rho_{\text{T}} \equiv \Delta\rho_2$, are related to the spintronic properties of the MI/NM bilayer via:

$$\Delta\rho_{\text{L}} \equiv \Delta\rho_1 = 2\,\rho_0\,\theta_{\text{SH}}^2\,\frac{\lambda}{t}\,Re\left[\frac{\lambda\,G_{\uparrow\downarrow}\,\tanh^2\left(\frac{t}{2\lambda}\right)}{\frac{1}{\rho_0} + 2\,\lambda\,G_{\uparrow\downarrow}\,\coth\left(\frac{t}{\lambda}\right)}\right] \quad (3)$$

$$\Delta\rho_{\text{T}} \equiv \Delta\rho_2 = -2\,\rho_0\,\theta_{\text{SH}}^2\,\frac{\lambda}{t}\,Im\left[\frac{\lambda\,G_{\uparrow\downarrow}\,\tanh^2\left(\frac{t}{2\lambda}\right)}{\frac{1}{\rho_0} + 2\,\lambda\,G_{\uparrow\downarrow}\,\coth\left(\frac{t}{\lambda}\right)}\right] \quad (4)$$

where $\lambda$ and $\theta$ are the spin diffusion length and the spin Hall angle of the NM layer, and $G_{\uparrow\downarrow}$ is the spin mixing conductivity of the MI/NM interface. Taking into account that $G_{\text{r}}/G_{\text{i}} \gg 1$ [11,33], we can combine Eqs. (3) and (4) to obtain

$$\frac{G_{\text{r}}}{G_{\text{i}}} \approx \frac{\Delta\rho_{\text{L}}}{\Delta\rho_{\text{T}}}\,\frac{1}{1 + 2\,\rho_0\,\lambda\,G_{\text{r}}\,\coth\left(\frac{t}{\lambda}\right)} \quad (5)$$

The SMR amplitude $\Delta\rho_{\text{L}}$ was evaluated in YIG/Pt and YIG/Au as shown in Figure 1(a,c,e) from longitudinal ADMR measurements at a field of 1 T to keep YIG saturated. $\Delta\rho_{\text{T}}$ was evaluated in the same samples from out-of-plane field Zero-Offset Hall measurements by driving the YIG film into magnetic saturation along the $z$-direction as shown in Figure 1(b,d,f).

For YIG/Pt, the longitudinal ADMR reveals the behavior expected from the SMR [Eq. (1)], namely a resistivity change of $\Delta\rho_{\text{L}}$ in the $\alpha$ and $\beta$ angles and no modulation in the $\gamma$ angle [Figure 1(c)]. The slight effect seen in $\gamma$ is consistent with a sample misalignment of less than 0.1°. From the measured resistivity $\rho_0 = 516$ n$\Omega$m in our Pt thin film, the values $\lambda_{\text{Pt}} = (1.2 \pm 0.2)$ nm and $\theta_{\text{Pt}} = 0.09 \pm 0.01$ can be inferred from the empirical relationships found for Pt thin films [37]. Together with the measured $\Delta\rho_{\text{L}} = (0.351 \pm 0.004)$ n$\Omega$m this leads to $(G_{\text{r}})_{\text{YIG/Pt}} = (3.8 \pm 1.0) \cdot 10^{14}\ \Omega^{-1}\ \text{m}^{-2}$ via Eq. (3). This value is in good agreement with previous reports in the same system [9,11,13,15,18,19]. When applying a magnetic field along the $z$-direction, the YIG film is gradually brought into saturation, while a proportional transverse magnetoresistance develops [Figure 1(d)]. The value $\Delta\rho_{\text{T}}$ corresponds to the saturation value of the resistivity change. Eq. (5) then yields $(G_{\text{r}}/G_{\text{i}})_{\text{YIG/Pt}} = 22 \pm 3$ for the YIG/Pt reference sample [Table 1]. This ratio is in good agreement with the theoretical calculations of $G_{\text{r}}/G_{\text{i}} \approx 20$ [33], as well as experimental values of $G_{\text{r}}/G_{\text{i}} = 16 \pm 4$ [18] and $G_{\text{r}}/G_{\text{i}} = 33$ [11].

The experimental values of $\Delta\rho_{\text{L,Au}} = (1.05 \pm 0.12)$ p$\Omega$m and $\Delta\rho_{\text{T,Au}} = (11.2 \pm 1.7)$ f$\Omega$m of YIG/Au are obtained in the same manner as for the YIG/Pt reference sample [see Figure 1(e,f)]. To actually measure a transverse magnetoresistance at this level of approximately 1 μ$\Omega$, special care must be taken to isolate the anomalous Hall signal from the much larger background contributions [inset in Figure 1(f)], which we accomplish by eliminating the longitudinal resistance by Zero-Offset Hall [36] and by accounting for the nonlinearity of the normal Hall effect itself [Suppl. Information, 4].

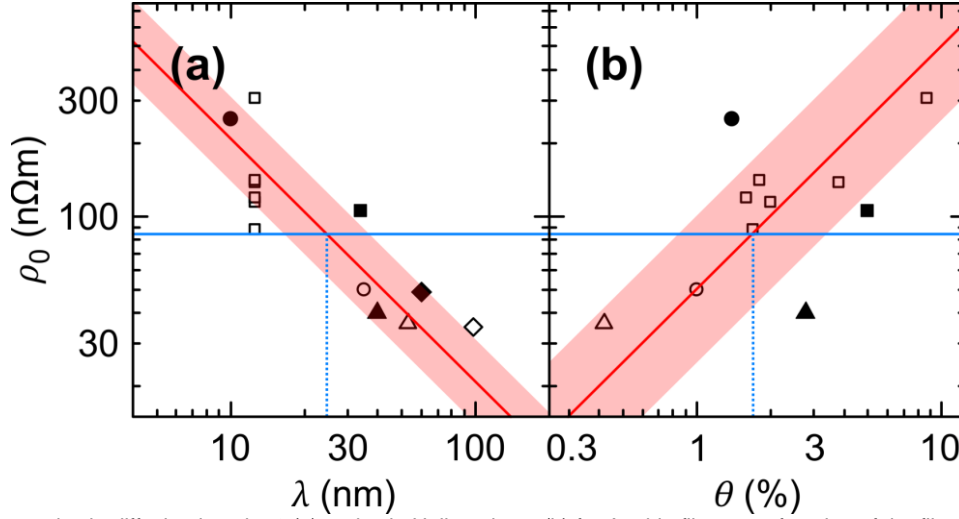

Figure 2: Reported spin diffusion lengths $\lambda$ (a) and spin Hall angles $\theta$ (b) for Au thin films as a function of the film resistivities $\rho_0$, and fits in context of the Elliott-Yafet relation (a) and resistivity dependence on the intrinsic spin Hall effect (b) shown as red lines with $1\sigma$ confidence bands. The data points are empirical data taken from: Kimura [38] (solid diamond), Brangham [39] (hollow squares), Vlaminck & Obstbaum [40,41] (hollow circle), Isasa [42] (hollow triangle), Niimi [43] (solid triangle), Mosendz & Obstbaum [44,45] (solid square), Laczkowski [46] (hollow diamond) and Qu [47] (solid circle). $\theta$ values from Refs. [42,43] have been multiplied by 2 for a proper comparison. The solid blue line denotes the resistivity of the Au film in the present YIG/Au system; dashed blue lines show the obtained spintronic quantities $\lambda$ and $\theta$ for our Au film.

In the following, we will provide an independent calculation of the SMR magnitudes for YIG/Au. First, we have to estimate the spintronic quantities $\lambda_{Au}$ and $\theta_{Au}$ of our Au film. Concerning $\lambda$, it is well established that the spin relaxation in metals is dominated by the Elliott-Yafet mechanism ($\lambda \propto \rho^{-1}$) [37,48–53]. Figure 2(a) illustrates how we obtain $\lambda_{Au} = (25^{+12}_{-8})$ nm based on the measured resistivity of the Au layer and empirical data [38–44,46,47]. This corresponds to a $\rho$-$\lambda$-product of $(2.1^{+1.0}_{-0.7}) 10^{-15}\,\Omega m^2$. Regarding $\theta$, different mechanism have been suggested to contribute in Au [39,42,43], but the origin of its SHE is not well established, yet. In order to estimate a reasonable value, we consider for simplicity that the intrinsic scattering contribution dominates the spin Hall angle. In this case, $\theta = \sigma_{SH}^{int} \times \rho$ holds, where the intrinsic spin Hall conductivity $\sigma_{SH}^{int}$ depends on the band structure and is thus constant for a given metal [21]. By taking reported data on Au [39–43,45–47], we estimate $\sigma_{SH}^{int} = (2.0^{+2.0}_{-1.0}) \left[\frac{\hbar}{2e}\right] 10^5\,\Omega^{-1}m^{-1}$, touching the upper end of theoretical predictions ranging from $(0.22\cdots0.9) \left[\frac{\hbar}{2e}\right] 10^5\,\Omega^{-1}m^{-1}$ [47,54,55]. The fitted $\sigma_{SH}^{int}$ value leads to $\theta_{Au} = 0.017^{+0.016}_{-0.008}$ for our Au film as shown in Figure 2(b).

In addition, we assume identical interface spin mixing conductivities in our reference Pt system and the Au system $G_{YIG/Pt} \equiv G_{YIG/Au}$, owing to the identical fabrication conditions, similar chemical qualities of the metals and similar Fermi energies and Sharvin conductivities of the metals [12]. Given these values, we can estimate the SMR magnitudes $\Delta\rho_{L,Au} = (0.7^{+2.0}_{-0.4})$ p$\Omega$m and $\Delta\rho_{T,Au} = (7^{+25}_{-5})$ f$\Omega$m for our YIG/Au sample. The calculated values are *quantitatively* consistent with the measured values. The uncertainty of the calculation will decrease in the future when the Elliott-Yafet scaling constant and intrinsic spin Hall conductivity are better understood for Au, as well as for other metals.

Table 1: Overview of the obtained quantities for YIG/Pt and YIG/Au studied here: Resistivity $\rho_0$, the relative longitudinal and transverse amplitudes of the spin Hall magnetoresistance $\Delta\rho_L$ and $\Delta\rho_T$, spin diffusion length $\lambda$, spin Hall angle $\theta$ and real and imaginary spin mixing conductivities $G_r$, $G_i$. * $\lambda$ and $\theta$ are derived from empirically observed scaling. ** Spin mixing conductivities are calculated for YIG/Pt and assumed to be identical for YIG/Au.

|  | Pt(2nm) | Au(10nm) |
|---|---|---|
| $\rho_0$ (n$\Omega$m) | 516 | 84.4 |
| $\Delta\rho_L/\rho_0$ | $6.8 \times 10^{-4}$ | $1.3 \times 10^{-5}$ |
| $\Delta\rho_T/\rho_0$ | $2.2 \times 10^{-5}$ | $1.3 \times 10^{-7}$ |
| $\lambda$ (nm) | $1.2 \pm 0.2$ * | $25^{+12}_{-8}$ * |
| $\theta$ | $0.09 \pm 0.01$ * | $0.017^{+0.016}_{-0.008}$ * |
| $G_r$ ($10^{14}\,\Omega^{-1}m^{-2}$) | $3.8 \pm 1.0$ | 3.8 ** |
| $G_r/G_i$ | $22 \pm 3$ | 22 ** |

We conclude that the observed magnitudes of the longitudinal and transverse magnetoresistance in our two systems, YIG/Au and YIG/Pt, are consistent with the same physical picture. Namely, the transverse magnetoresistance in these MI/NM systems can be fully understood as emergent from the transverse part of the spin Hall magnetoresistance due to the imaginary component of the spin-mixing conductivity. No evidence points to a contribution of proximity magnetization via the anomalous Hall effect. Au is prototypical for materials with a low resistivity and intermediate spin Hall angle, which indicates that the conventional SMR theory applies in this regime. In addition, the possibility of measuring both the longitudinal and the transverse spin Hall magnetoresistance amplitudes for a wide range of MI/NM interfaces provides an elegant way to study the spin-mixing conductivity and its fundamental dependencies.

## Supplementary Material

Supplementary material contains further details on the transport measurements including the approach to compensate for the normal Hall effect and remark on the absence of the planar Hall effect in Zero-Offset Hall measurements. Electron microscopy imaging of the samples in cross section is reported as well. Additional measurements are reported to assess the contribution of the anomalous Hall effect due to the thermal spin pumping. The data includes hysteresis loops measured at different current densities as well as transport data taken at higher harmonics.

## Acknowledgements

Support by the Structural Characterization Facilities Rossendorf at the Ion Beam Center (IBC) at the HZDR is greatly appreciated. The work was financially supported in part via the German Research Foundation (DFG) Grant MA 5144/9-1, the BMBF project GUC-LSE (federal research funding of Germany FKZ: 01DK17007), the BMWi project WiTenso (03THW12G01), by the Spanish MINECO under the Maria de Maeztu Units of Excellence Programme (MDM-2016-0618) and under Project No. MAT2015-65159-R and by the Regional Council of Gipuzkoa (Project No. 100/16). J.M.G.-P. thanks the Spanish MINECO for a Ph.D. fellowship (Grant No. BES-2016-077301).

# Supplementary Information

## 1. Thermal spin pumping induced anomalous Hall effect

It has been reported that thermal gradients can lead to thermal spin pumping, which in turn can lead to proximity magnetization of Au thin films on YIG depending on the strength of the thermal gradient [1]. Such proximity magnetization will also lead to an anomalous Hall effect but would be not due to the spin Hall magnetoresistance. We argue that the contribution of thermal spin pumping induced proximity magnetization is negligible in our present study.

Several references and statements within D. Hou et al. [1] confirm that Au thin films are not expected to reveal any static magnetic proximity effect. This is also exemplified by the experiments in that work, which were performed without a thermal gradient resulting in no detected AHE signal. The authors argue that thermal spin pumping leads to magnetization in Au. As our Au film is Joule heated due to transport experiments, we expect a thermal gradient from Au towards YIG also in our samples.

D. Hou et al. used an inverted notation of Hall voltage which is evident from their positive slope of the normal Hall effect in their Figure 2(b). Therefore, the positive slope anomalous Hall effect (AHE) curves in that work are identically signed as our negative slope AHE curves. Taking into account our twice lower Au thickness, we would be able to explain our AHE signal magnitude with a purported thermal gradient of about 3.5 K/mm. If this thermal gradient is indeed present and is the origin of our Hall signal, the expectation from the cited work would be an increase in the magnitude of AHE curves when larger currents are used to probe the Au film.

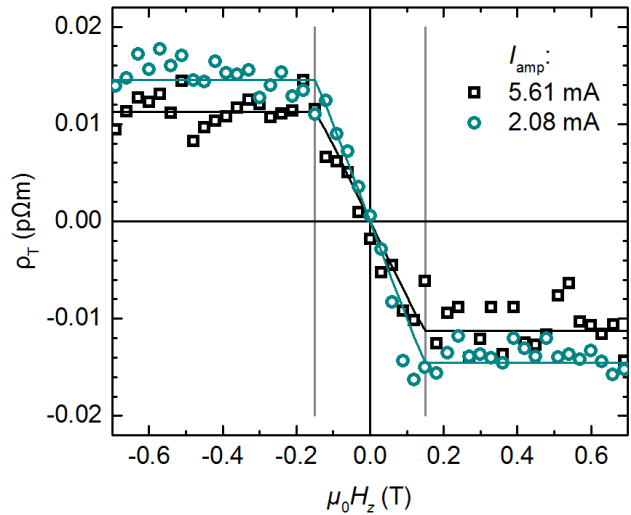

Figure S1: Variation of the transverse magnetoresistance upon employing different probe current amplitudes $I_{amp}$.

We did measure our sample at various levels of current. Due to our narrow stripe pattern these currents lead to varying extents of Joule heating and thus thermal gradients orthogonal to the film plan. We did not observe an increasing trend of the AHE magnitude with increasing probe current. In fact, despite a more than 7-fold variation of the Joule heating dissipation, we did not observe any significant change in the magnitude of the AHE effect [Figure S1]. The slight (but non-significant) variation of the AHE magnitude is due to the processing of the nonlinear background, which is slightly different scan to scan due to long term nature of the scans. Even if the variation of the AHE magnitude with the current amplitude is taken to be real, it is opposite to that expected from thermal spin pumping described by D. Hou et al. Therefore, we conclude that the thermal spin pumping induced AHE is negligible for the total magnitude of our observed transversal magnetoresistance.

Furthermore, it is also possible to separate the contributions from the transverse spin Hall magnetoresistance and from thermal spin pumping AHE to the total transverse magnetoresistance through harmonic measurements. The signal in Figure S1 follows from the transverse voltage at the excitation current frequency (first harmonic) and can contain contributions from both effects. In contrast, the voltage at the third harmonic frequency should contain only contributions from thermal spin pumping: When we supply a low frequency ($f$) AC current to probe our Au films, we expect to heat the Au film not only to an equilibrium value – which we discussed above – but the Au film would also exhibit temperature variations with a $2f$ periodicity. The temperature would peak every time the probe current sine wave is at its tips and temperature would dip below the average value at the zero-crossings of the sine wave. According to D. Hou et al. such a temperature behavior would result in more transient magnetization being present in Au while the probe current sine wave is near its tips. Therefore, the resulting AHE voltage would be larger than a pure sine wave near the tips of the sine wave probe current resulting in an additional third harmonic Hall signal. This third harmonic Hall signal should exhibit the same behavior as the magnetization of YIG, i.e. an antisymmetrical curve saturating at about 0.15 T.

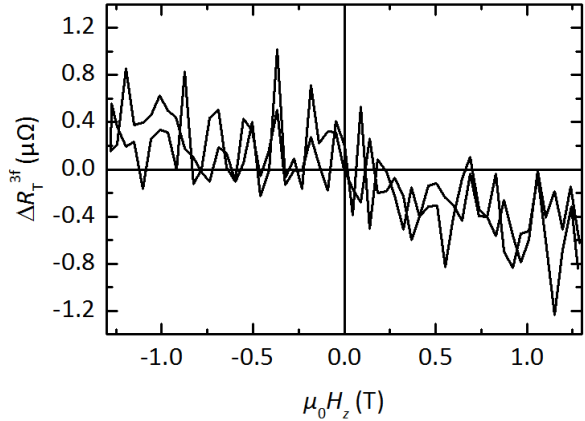

Figure S2: Third harmonic Hall signal using $I_{\text{amp}} = 5.61$ mA.

We indeed also probed this third harmonic signal and found no significant YIG-like variation of it as we swept the field [Figure S2]. Fitting a YIG like signal to the third harmonic Hall signal of Figure S2 leads to a thermal spin pumping AHE of about $2$ fΩm. This is a) much lower than the first harmonic transverse magnetoresistance magnitude we observe and b) is questionable to exist at all in light of the data in Figure S2 which could be also just a linear trend due to the normal Hall effect and slightly non-sinusoidal excitation.

Judging from the signal levels reported by D. Hou et al., we should have definitely observed these trends using our very sensitive methodology if the thermal gradient was sufficient and the effect as strong as reported. We hypothesize that the absence of the thermal gradient AHE signals in our study could be due to either a lower thermal gradient or the different preparation conditions. Regarding the latter, although D. Hou et al. did not provide structural information on their YIG/Au interface, we believe that it is of more coherent quality due to their etching and in-situ-prenannealing procedure prior to Au deposition. Instead, to observe the SMR of Au/YIG, no interface optimization was found necessary. As a result, we deposited Au on untreated YIG substrates, yielding dense continuous films, but no epitaxy probably because of amorphous surface termination of the YIG surface due to adsorbates. These adsorbates could also render the thermal spin pumping inefficient in our samples.

2. **Sample preparation and measurement details**

The reason we chose 10 nm Au as our main sample is because we wanted to avoid discontinuous metal films. Such discontinuities appear due to the bad wetting of the noble metals on the untreated YIG, especially for metals with low melting temperatures, such as Cu, Ag, Au [Figure S3]. Discontinuous layers can give rise to spurious effects like wrong resistivity values when assuming nominal thicknesses which would be problematic for our determination of the spin Hall angle and the spin diffusion length.

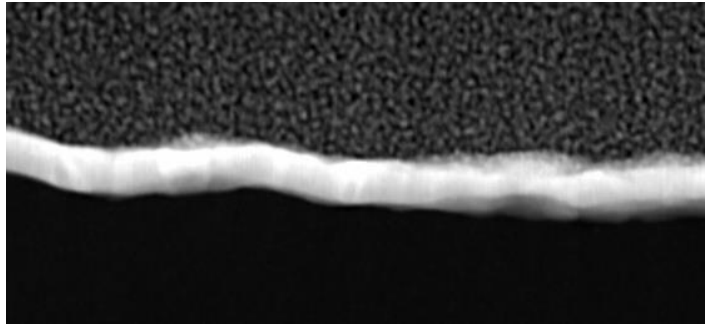

Figure S3: Cross-sectional STEM image of 10 nm film of Au on an untreated liquid phase epitaxy YIG-on-GGG substrate. The Au film is fully continuous and has a homogenous thickness.

On the other hand, films thicker than 10 nm would produce so much shunting of the interfacial SMR effect (or magnetic proximity effect), that the detection of the transversal signal would be no longer possible. We want to stress that the presented measurements have been very challenging even in this 10 nm Au sample. Even though we used a precisely patterned Hall cross with only about 0.3 % contact skew, the residual longitudinal resistance background amounted to roughly $11\ \mathrm{m\Omega}$, which is roughly $10^5$ over the minimum necessary precision to detect the AHE curves of about $100\ \mathrm{n\Omega}$ and places high demands for continuous dynamic measurement range. In addition, to achieve this precision at a noise density of about $2\ \mathrm{\mu\Omega}/\sqrt{\mathrm{Hz}}$ (at 2 mA current amplitude) required approximately 400 seconds of integration for each of the 100 field bins, amounting to an integration time of approximately half a day. Therefore, although our purpose made measurement device unites great continuous dynamic range and extremely low noise, we had to employ sophisticated drift compensation methods to achieve a low enough corner frequency to actually resolve the AHE effect in our samples.

### 3. Planar Hall effect

When measuring the transverse voltage drop developed by a slab of anisotropically conducting material, a planar Hall effect is generally observed. In context of polycrystalline metal films used in spintronics, such anisotropy commonly appears as a result of e.g. the anisotropic magnetoresistance [2,3] or spin Hall magnetoresistance [4]. When a voltage is applied to this anisotropically conducting thin film, the current will in general not flow collinearly with the voltage gradient, but experience a transverse deflection towards the high conductivity axis, which results in an equilibrium transverse voltage drop. In a stationary measurement layout, this planar Hall effect only vanishes for a) perfectly isotropic conduction or b) conduction exactly along the high or low conductivity axes.

However, despite its name, the planar Hall effect is fundamentally different from other Hall effects such as the normal and anomalous Hall effects (real Hall effects). When averaging all in-plane current directions, the planar Hall effect assumes both positive and negative values as current experience alternatingly left-handed and right-handed deflection. In contrast, the real Hall effects persist as they experience the same handedness of deflection for any in-plane current direction. Therefore, planar Hall effects disappear in Zero-Offset Hall measurements [5], while the real Hall effects are maximally preserved.

### 4. Normal Hall effect

In order to study tiny anomalous Hall-like signals, it is mandatory to dynamically reject the influence of the longitudinal resistance, which – even for carefully patterned – Hall cross structures can be much larger than the Hall signal of interest [5]. While this is elegantly achieved using the Zero-Offset Hall measurement scheme, the normal Hall effect cannot be rejected using this approach. D. Hou et al. reported an approach to reject the normal Hall effect via a lock-in method [1], but this is possible only when the origin of the anomalous Hall effect can be switched on and off, which is not the case in our sample.

We remove the normal Hall effect after the measurement by subtracting it from the measured transverse resistance. The observed normal Hall effect is not completely linear in the magnetic field for two reasons: a) our magnetic field reading is not fully linear over the actual magnetic field mainly due to nonlinearities of the Si hall probe. b) the normal Hall effect of the thin metal (Pt or Au) films can be slightly non-linear itself. These influences cause a total nonlinearity of the normal Hall effect in the range of $10^{-3}$ to $10^{-2}$, which is non-negligible in our study. The function we use to model the normal Hall effect is composed of three contributions:

$$\rho_{\mathrm{NHE}} = A_1 H_z + A_2 \left(1 - \mathrm{sech}\left(\frac{H_z}{H_{\mathrm{NL}}}\right)\right) + A_3 \tanh\left(\frac{H_z}{H_{\mathrm{NL}}}\right) \qquad (1)$$

where $H_z$ is the magnetic field applied out-of-plane, $H_{\mathrm{NL}}$ is a characteristic field determining the shape of the nonlinearity and $A_{1,2,3}$ are scaling constants. The first contribution $A_1 H_z$ is the largest by far, while the sech and tanh functions capture smaller even and odd nonlinearities, respectively. One motivation for this model choice is that it cannot accidently introduce YIG-like signals, when $H_{\mathrm{NL}}$ is sufficiently large, namely $H_{\mathrm{NL}} \gtrsim 2 H_{\mathrm{sat,YIG}}$ with $\mu_0 H_{\mathrm{sat,YIG}} \approx 0.16\,\mathrm{T}$. When this is fulfilled, the tanh contribution is essentially linear in the relevant field range. For the presented measurements, we used $\mu_0 H_{\mathrm{NL}} \approx 1.1\,\mathrm{T}$ to model and subtract the normal Hall contribution to $\rho_{\mathrm{T}}$ [Figure 1(d,f) of the main text].